\documentclass{ieeeaccess}
\usepackage{cite}
\usepackage{url,amsfonts,amsmath,graphicx,enumitem,cite, hyperref}
\usepackage{amsmath,amssymb,amsfonts}
\usepackage{algorithmic}
\usepackage{graphicx}
\usepackage{textcomp}
\usepackage{tabularx,booktabs}
\usepackage{multirow}
\usepackage{color}

\def\BibTeX{{\rm B\kern-.05em{\sc i\kern-.025em b}\kern-.08em
    T\kern-.1667em\lower.7ex\hbox{E}\kern-.125emX}}
\begin{document}
\history{Date of publication xxxx 00, 0000, date of current version xxxx 00, 0000.}
\doi{10.1109/ACCESS.2017.DOI}

\title{CITISEN: A Deep Learning-Based Speech Signal-Processing Mobile Application}
\author{
\uppercase{Yu-Wen Chen},
\uppercase{Kuo-Hsuan Hung},
\uppercase{You-Jin Li},
\uppercase{Alexander Chao-Fu Kang},
\uppercase{Ya-Hsin Lai},
\uppercase{Kai-Chun Liu},
\uppercase{Sze-Wei Fu},
\uppercase{Syu-Siang Wang},
\uppercase{Yu Tsao}
\IEEEmembership{Senior Member, IEEE}}
\address {Research Center for Information Technology Innovation at Academia Sinica, Taipei, Taiwan}




\begin{abstract}
This study presents a deep learning-based speech signal-processing mobile application known as CITISEN. The CITISEN can perform three functions: speech enhancement (SE), model adaptation (MA), and background noise conversion (BNC), which allow CITISEN to be used as a platform for utilizing and evaluating SE models and flexibly extend the models to address various noise environments and users. For SE, CITISEN downloads pretrained SE models on the cloud server and then uses these models to effectively reduce noise components from prerecording or instant recording provided by users. When it encounters noisy speech signals with unknown speakers or noise types, the MA function allows CITISEN to improve the SE performance effectively. A few audio files of unseen speakers or noise types are recorded and uploaded to the cloud server and then used to adapt the pretrained SE model. Finally, for BNC, CITISEN removes the original background noise using an SE model and then mixes the processed speech signal with new background noise. The novel BNC function can evaluate SE performance under specific conditions, cover people’s tracks, and provide entertainment. The experimental results confirmed the effectiveness of SE, MA, and BNC functions. Compared with the noisy speech signals, the enhanced speech signals by SE achieved about 6\% and 33\% of improvements, respectively, in terms of short-time objective intelligibility (STOI) and perceptual evaluation of speech quality (PESQ). With MA, the STOI and PESQ could be further improved by approximately 6\% and 11\%, respectively. Note that the SE model and MA method are not limited to the ones described in this study and can be replaced with any SE model and MA method. Finally, the BNC experiment results indicated that the speech signals of original and converted backgrounds have a close scene identification accuracy and similar embeddings in an acoustic scene classification model. Therefore, the proposed BNC can effectively convert the background noise of a speech signal and be a data augmentation method when clean speech signals are unavailable.

\end{abstract}

\begin{keywords}
speech enhancement, model adaptation, background noise conversion, deep learning, mobile application.
\end{keywords}

\titlepgskip=-15pt

\maketitle

\section{Introduction}
\label{sec:introduction}

In recent years, a wide variety of speech-related applications have been developed. Most of these applications are highly convenient for human–human and human–machine communications. However, the following long-existing and critical issues that may limit the achievable performance of these applications remain to be solved: speech distortions caused by additive/convolutional noises and channel/device effects \cite{varga1993assessment, giraud1997auditory, kinoshita2013reverb, beutelmann2006prediction, steeneken1980physical, sankar1996maximum}. Identifying an effective method of addressing this distortion issue is a critical and challenging task, and numerous approaches have been proposed to this end, among which speech enhancement (SE) is notable.

The goal of SE is to transform noisy speech signal into enhanced speech signal with improved quality and intelligibility \cite{loizou2013speech, benesty2005speech}. In the past several decades, SE has been widely used as a front-end unit in many voice-based applications, such as automatic speech recognition \cite{li2015robust}, speaker identification \cite{el2007evaluation}, speech coding \cite{li2011comparative}, hearing aids \cite{levit2001noise}, and cochlear implants \cite{lai2016deep, chen2015evaluation}. The existing SE methods can be divided into two classes. In the first class, SE methods design a filter or function to attenuate noise components. Examples of methods in this class include the Wiener filter and its extensions \cite{scalart1996speech, hansler2006topics, chen2008fundamentals, li2011two}, the minimum mean square error spectral estimator (MMSE) \cite{ ephraim1985speech, mcaulay1986speech, quatieri1992shape}, Karhunen-Loeve transform \cite{mittal2000signal}, maximum a posteriori spectral amplitude estimator \cite{ephraim1992statistical, suhadi2010data, su2013speech}, maximum likelihood spectral amplitude estimator \cite{kjems2012maximum,mcaulay1980speech}, linear prediction models \cite{makhoul1975linear}, orthogonal polynomial-based method \cite{jassim2014enhancing}, super-Gaussian-based methods \cite{ lotter2005speech , zou2007speech}, and the hybrid of orthogonal polynomial and super-Gaussian \cite{mahmmod2019speech}. Most SE methods of the first class have a common limitation: the inability to effectively contrast non-stationary noise signals in real-world scenarios under unexpected acoustic conditions.
 
SE methods in the second class are based on machine-learning algorithms; these methods typically prepare a model for noisy-to-clean transformation in a data-driven manner. Notable SE methods belonging to this class include hidden Markov models \cite{rabiner1986introduction}, non-negative matrix factorization \cite{NIPS2000_f9d11525, wilson2008speech, mohammadiha2013supervised}, compressive sensing \cite{wang2016compressive}, and robust principal component analysis \cite{wu2017wavelet}. In addition, artificial neural networks (ANNs), as a successful machine-learning model, have been used for SE because of their powerful nonlinear transformation capability. In \cite{tamura1989analysis,xie1994family,wan1999networks, tchorz2003snr}, a shallow ANN was used to map noisy speech signals to clean ones. More recently, various types of ANNs with deep structures have been used for SE (e.g., deep neural networks (DNNs) \cite{ vincent2010stacked, wang2013towards, lu2013speech, xu2014regression,liu2014experiments, abdullah2021towards, siniscalchi2021vector}, deep recurrent neural networks and long-short term memory (LSTM) networks \cite{wollmer2013feature, weninger2015speech, shan2018novel, saleem2020learning},  convolutional neural networks (CNNs) \cite{fu2017raw, fu2018end, pandey2019new}, and convolutional recurrent neural networks (CRNNs) \cite{tan2018convolutional, hu2020dccrn}). Also, \cite{qi2020exploring} proposed a hybrid architecture of CNN and a tensor-train layer and compared the performance between DNN and CNN.

To improve the performance of these ANN-related approaches, several SE studies have applied a generative adversarial network (GAN) model \cite{li2018conditional, phan2020improving, pascual2017segan, yang2020improving}. The GAN model is used to generate enhanced samples for a discriminator to determine whether the input follows the distribution of a real clean speech signal. In addition, some researchers applied a transformer technique to perform SE, in which the attention mechanism was utilized to capture long-term temporal correlations to extract clean components from noisy input \cite{tang2020joint, li2020noise, kim2020t, liang2020real, lan2020embedding}. Moreover, instead of using a large amount of training data to perform SE, a transfer learning technique has been commonly used to enhance the generalization of models in unseen environments. For example, \cite{pascual2018language} fine-tuned the generator in a pretrained GAN-based SE model with small amounts of data and confirmed the efficiency of transfer learning. In \cite{wang2020cross}, the authors proposed the use of a teacher-student learning strategy to adapt an SE model to unlabeled noisy speech signals. Furthermore, the FA-MK-MMD approach was proposed in \cite{liang2020transfer} to train a neural network model from the labeled source domain to extract the shared representation to enhance the unlabeled input. Although the effectiveness of these SE approaches has been verified, their performance in mobile applications is yet to be confirmed. 

In this study, we present a speech signal processing mobile application called CITISEN\footnote{\textcolor{blue}{CITISEN GitHub Page: \url{https://github.com/yuwchen/CITISEN}} If there is any problem with testing pretrained SE models on CITISEN, please contact the corresponding author.}. CITISEN is a standardized SE software with a user interface that can be used as a platform for utilizing and evaluating newly performed deep-learning-SE models by simply replacing the default settings with the associated model. Based on SE, two extended functions—model adaptation (MA) and background noise conversion (BNC)—were also implemented in CITISEN. The MA function was built to further improve the SE performance for a specific user or under certain noise environments. The adaptation data were prepared by the users to meet their requirements, thus making the framework a customized tool. The BNC function converts the original background noise to another one. BNC can be used to evaluate SE performance under practical conditions. In this condition, the residual noises in an enhanced source speech signal are combined with different background interference and affect the quality and intelligibility of a target speech signal. In addition, the BNC can be used to cover people’s tracks by converting the original environment noises to noises from other places when a positioning system is unavailable or not being used because of limited access to the technology or the lack of intention. Furthermore, the BNC can also be used for entertainment purposes, such as adding background music or sound effects.

The contribution of this study is summarized as follows:
\begin{itemize}

\item To the best of our knowledge, the proposed CITISEN is the first to integrate BNC and MA functions with SE in a mobile application. 

\item CITISEN has a user interface for performing SE on an prerecording or instant recording. The experimental results confirmed the SE function of improving short-time objective intelligibility (STOI) \cite{taal2011algorithm} and perceptual evaluation of speech quality (PESQ) \cite{rix2001perceptual} scores. 

\item CITISEN has an MA function that allows users to adapt the SE models to unseen background noises or speakers. The MA function is proven to provide notable STOI and PESQ improvements compared to the results without MA. 

\item CITISEN provides a novel BNC function that can evaluate SE performance under specific conditions, cover people’s tracks, and provide entertainment. The listening test results indicated that the BNC function could convert the background noise while maintaining the clarity and intelligibility of the converted speech signals. 

\item An acoustic scene classification (ASC) model was used to evaluate the BNC performance. The results showed that new background noise could be successfully recognized. Moreover, the ASC embeddings suggested that the conversation results from a silent background were close to a noisy background. Therefore, the BNC function can potentially serve as a data augmentation method for the ASC model when clean speech signals are unavailable. 

\item By simply replacing the settings with the associated model, CITISEN can utilize and evaluate other deep learning-based SE models not described in this study. Therefore, CITISEN can effectively reduce the development interval for converting SE models to industrial applications. 
\end{itemize}

The remainder of this paper is organized as follows. Section II reviews related works. Section III elaborates the functions and user interface of CITISEN. Section IV presents the experimental setup and results. Finally, Section V provides some concluding remarks regarding this research.

\section{Related works}
In this section, we first review one traditional filter-based SE method and four neural-network-based SE models used for comparison in the experiments. Then, we introduce the concept of MA.

\subsection{Traditional gain function-based SE method}

In the SE task, we generally assume that the noisy speech signal $x$ contains a clean speech signal $s$ and noise signal $v$.

\begin{equation} \label{eq:SE_defination}
x=s+v
\end{equation}

\begin{figure}[htbp!]
{\centering \includegraphics[width=0.8\columnwidth]{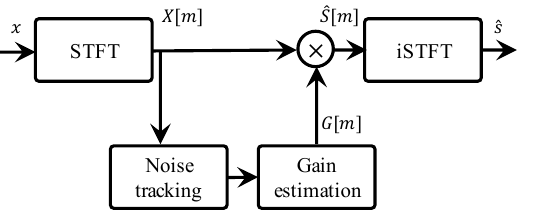}
\caption{Traditional filter-based SE architecture. STFT and iSTFT denote the short-time Fourier transform and inverse STFT, respectively.}\label{fig:mmse}}
\end{figure}

For the MMSE SE \cite{ephraim1985speech, tsao2016generalized} approach, the time-domain signal, $x$, is first converted to a spectral feature, $X$, using the short-time Fourier transform (STFT). After the STFT, Eq. \ref{eq:SE_defination} can be expressed as:

\begin{equation}
X[m]=S[m]+V[m]
\end{equation}

where $m$ denotes the $m$th frequency bin in the entire set of spectral features. By estimating the a priori and a posteriori signal-to-noise ratio (SNR) statistics based on a noise-estimation approach, we can estimate a function $G[m]$. The enhanced speech signal, $\hat{S}[m]$, is obtained by filtering $X[m]$ through $G[m]$. Finally, an inverse STFT (iSTFT) is applied to convert the spectral features $\hat{S}$ to the time-domain signal $\hat{s}$, as shown in Fig.\ref{fig:mmse}.

\subsection{Neural-network-based SE method}

In this work, we used one waveform-based SE model, fully convolutional network (FCN) \cite{fu2017raw}, and three spectral-based SE models, namely, deep denoising autoencoder (DDAE) \cite{lu2013speech}, LSTM\cite{weninger2015speech}, and CRNN\cite{tan2018convolutional}. Table. \ref{tab:model_summary} summarizes the NN models used in this study. Similar to traditional SE methods, the goal of the neural-network-based SE is to find the enhanced speech signal $\hat{s}$ that is close to the clean speech signal ${s}$.

\begin{table*}
\begin{center}
\begin{tabular}{c|cccl|c} 
\hline\hline
\multicolumn{1}{l|}{} & \multicolumn{4}{c|}{Models used in Section IV~} & \multirow{2}{*}{Relevant Works}  \\ 
\cline{2-5}\multicolumn{1}{l|}{} & Ref. & Feature type & NN layer & Highlight & \\ 
\hline\hline
FCN-based & \cite{fu2017raw} &W& Conv& 
\begin{tabular}[t]{@{}l@{}}The input length can be varied, and the local feature\\ structures can be preserved.\end{tabular}
& \cite{fu2018end, pandey2019new, phan2020improving}\\
\hline
DDAE-based& \cite{lu2013speech} & S& Dense & The model structure is simple.& \cite{vincent2010stacked, xu2014regression,liu2014experiments, abdullah2021towards} \\
\hline
LSTM-based& \cite{weninger2015speech} & S& \begin{tabular}[t]{@{}c@{}}LSTM\\Dense\end{tabular}& LSTM is known for capturing temporal features. & \cite{wollmer2013feature, shan2018novel, saleem2020learning, liang2020real} \\
\hline
CRNN-based & \cite{tan2018convolutional} & S            & \begin{tabular}[t]{@{}c@{}c@{}}Conv\\LSTM\\Dense\end{tabular}    & 
\begin{tabular}[t]{@{}l@{}}The CRNN-based model takes advantage of both\\the
convolutional and LSTM layers.\end{tabular}& \cite{lan2020embedding, hu2020dccrn} \\
\hline\hline
\end{tabular}
\end{center}
 \caption{Summary of NN models used in this study. The \textbf{W} and \textbf{S} in feature type column represent waveform-based and spectral-based input, respectively. \textbf{Conv} is the abbreviation of convolution. }\label{tab:model_summary}
\end{table*}

\subsubsection{FCN-based SE model}
\begin{figure}[htbp!]
{\centering \includegraphics[width=0.75\columnwidth]{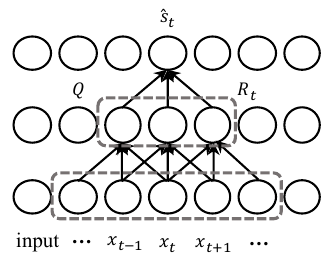}
\caption{FCN-based SE architecture.}\label{fig:fcn}}
\end{figure}

Fig. \ref{fig:fcn} shows an FCN model, which is similar to a conventional CNN, but all the fully connected layers are removed. As reported in \cite{fu2018end}, the FCN model can address the high-and low-frequency components of the raw waveform simultaneously. The relation between the output sample $\widehat{{s}}_{t}$ and the connected hidden nodes ${R}_{t}$ can be represented by:

\begin{equation}
    \widehat{s}_{t} = Q^TR_{t}
\end{equation}

where $Q$ denotes one of the learned filters and subscript $t$ indexes the time step. Then, the objective function of the FCN-based SE model is defined as:

\begin{equation}
\mathcal{L} ( \theta_{F} ) = ||\widehat {s} -s||^2
\end{equation}

where $\theta_{F}$ denotes the model parameters of FCN.

\subsubsection{DDAE-based SE model}

During the training of DDAE, noisy-clean speech signal pairs were used to compute the mapping function from noisy to clean spectral (logarithm amplitude in this study) features. The DDAE model aims to transform the noisy speech signal to a clean speech signal by minimizing the reconstruction error between the predicted $\widehat{{S}}$ and the reference clean spectral features ${S}$ such that:

\begin{equation}
\theta^*_{D} = \underset{\theta_{D}}{\arg\min}  \mathcal{L}(\theta_{D}) + \rho C(\theta_{D}),
\end{equation}
with
\begin{equation}
\mathcal{L}(\theta_{D}) = \|\widehat{S}-{S}\|^2,
\end{equation}

where $\theta_{D}$ denotes the model parameters of DDAE. $\rho$ is a constant that controls the trade-off between the reconstruction accuracy and regularization term $C(\theta_{D})$ \cite{lu2013speech} and is determined through the validation set in the training process. In this study, to simplify and compare with other methods, we set $\rho$ to $0$.

Given noisy spectral features ${X}$, the DDAE estimates clean speech signal by:
\begin{equation}
\begin{array}{c}
h_1({X})= \sigma({W}_1{X} + b_1),\\
\vdots\\
h_{D-1}(X)= \sigma (W_{L-1} h_{L-2} (X)+b_{L-1}),\\ 
\widehat{S}= W_L h_{L-1} (X)+b_L,
\end{array}
\end{equation}

where ${W_1 . . . W_L}$ and ${b_1 . . .b_L}$ are the weight matrices and bias vectors, respectively, and $L$ is the number of layers. In addition, $\sigma$ is a vector-wise non-linear activation function sigmoid. 

\subsubsection{LSTM-based SE model}

Because LSTM can capture the temporal relation of speech signal, it has proven to deliver promising results in SE \cite{weninger2015speech}. The objective function of the LSTM-based SE model is close to that of the DDAE model, which is to find the best model parameters of LSTM $\theta_{L}$ that can minimize:

\begin{equation}
\mathcal{L}(\theta_{L}) = \|\widehat{S}-{S}\|^2,
\end{equation}
In this study, we used the LSTM unit defined as follows:
\begin{equation}
\begin{array}{c}

i_n = \sigma(W_i X_n + U_i h_{n-1} + b_i),\\
o_n = \sigma(W_o X_n + U_o h_{n-1} + b_o),\\
f_n = \sigma(W_f X_n + U_f h_{n-1} + b_f),\\
g_n = \tanh(W_g X_n + U_g h_{n-1} + b_g),\\
c_n = f_n\odot c_{n-1} + i_n\odot g_n,\\
h_n = o_n \odot \tanh(c_n)

\end{array}
\end{equation}

where $X_n$, $f_n$, $i_n$, $o_n$, $g_n$, $c_n$, and $h_n$ represent the input, forget gate, input gate, output gate, cell input activation, cell state, and hidden state vectors, respectively, and the subscript $n$ indexes the frame step. In addition, $W_q$ and $b_q$ denote the weights and biases, respectively, where the subscript $_{q}$ can either be the input gate $i$, output gate $o$, forget gate $f$, or memory cell $g$, and $\odot$ represents element-wise multiplication.

\subsubsection{CRNN-based SE model} 
\begin{figure}[htbp!]
{\centering \includegraphics[width=0.75\columnwidth]{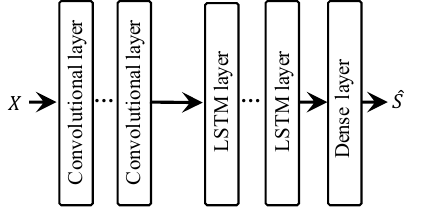}
\caption{CRNN-based SE architecture.}\label{fig:crnn}}
\end{figure}

The CRNN in this work combines a CNN, LSTM, and Dense layer. Previous work indicated that CRNN could lead to better objective intelligibility and perceptual quality than an LSTM model with fewer trainable parameters \cite{tan2018convolutional}. The architecture of the CRNN-based SE model is shown in Fig. \ref{fig:crnn}.

\subsection{Model adaptation} 
When operating SE in a real-world scenario, unknown noise types and new users are often encountered. Therefore, in many cases, the testing data may not be adequately covered by the trained SE model. Such training/testing mismatches in acoustic characteristics may considerably degrade SE performance. To effectively address this mismatch issue, an adaptation of an SE model is required. Thus far, various MA approaches have been proposed \cite{chopra2013dlid, fine-tune1, ganin2016domain, finn2017model}. The main concept of MA is to adjust the parameters of a pretrained model (prepared using training data) based on a small set of testing data to reduce the influence of training/testing mismatches. Because the adapted SE models match the testing conditions, the SE performance can be improved.

\section{The presented CITISEN Application}

\begin{figure}[htbp!]
{\centering \includegraphics[width=1.0\columnwidth]{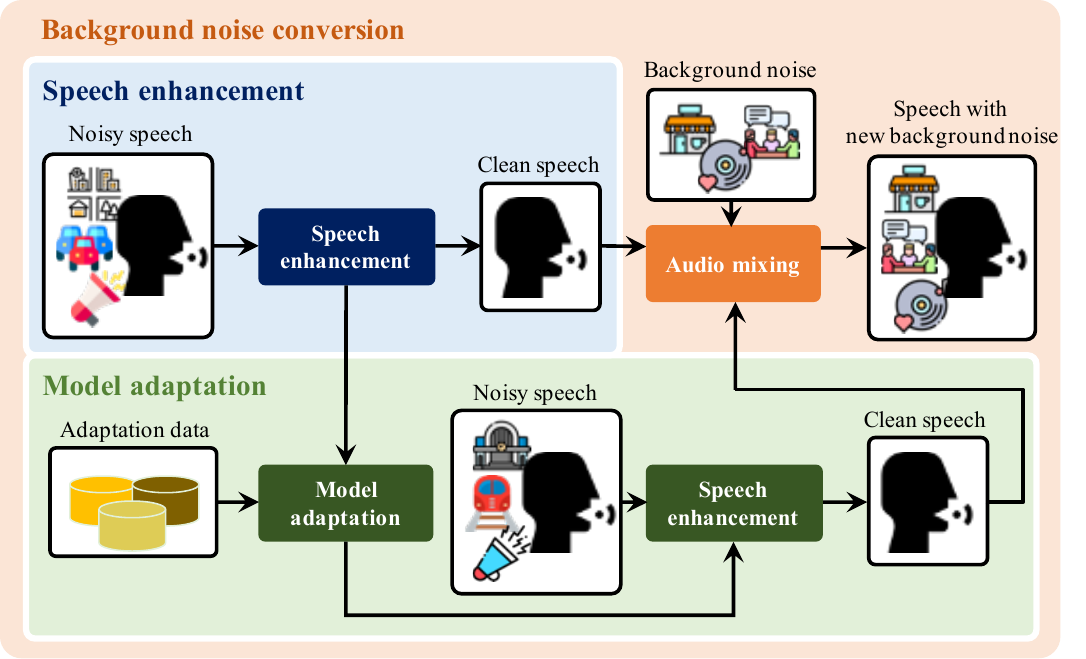}
\caption{The SE, BNC, and MA functions in CITISEN}\label{fig:function_intro}}
\end{figure}

CITISEN has three functions, including SE, MA, and BNC. For SE, CITISEN can enhance the quality and intelligibility of noise signals by reducing noise components from the speech signals. Then, for MA, CITISEN can further improve the results of SE by fine-tuning the SE model with uploaded data. Finally, for BNC, CITISEN can replace the original background noise with specified background noise. The functions of CITISEN are illustrated in Fig. \ref{fig:function_intro}.

\begin{figure*}
\begin{center}\includegraphics[scale=0.8]{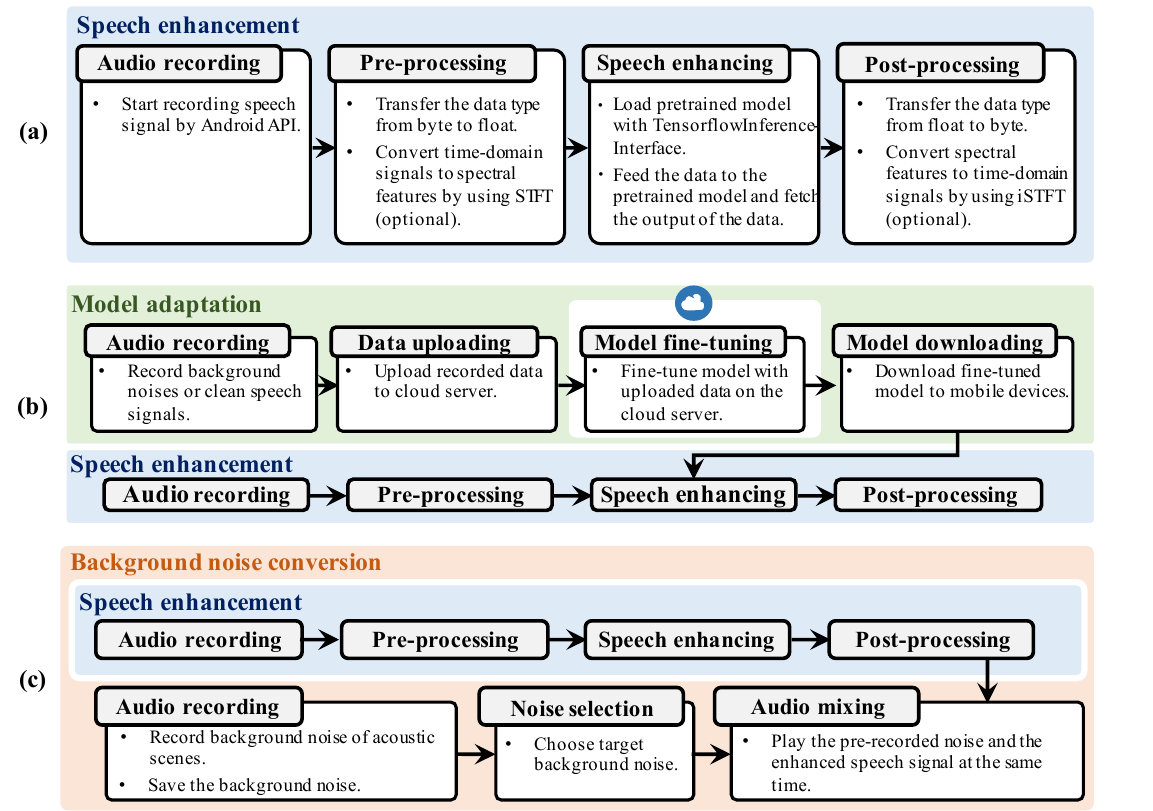}
\caption{Implementation details of three functions in CITISEN. Figure (a), (b), and (c) are SE, MA, BNC functions, respectively. For SE, CITISEN downloads pretrained SE models on the cloud server and then uses them to enhance prerecording or instant recording provided by users. For MA, a few audio files of unseen speakers or noise types are recorded and uploaded to the cloud server then used to adapt the pretrained SE model. For BNC, CITISEN removes the original background noise using an SE model, then mixes the processed speech signal with new background noise.}\label{fig:implementation}
\end{center}
\end{figure*}

\subsection{SE Function}

SE is a major function of CITISEN. As shown by the blue block in Fig. \ref{fig:function_intro}, given the noisy speech signal, the SE function removes background noises and generates the enhanced speech signal with improved quality and intelligibility. The SE models were trained in a cloud server, and the trained models were loaded into mobile devices. Because the model is trained and saved in a cloud server, mobile devices do not need to have a huge computational resource. When connected to the Internet, mobile devices automatically download updated SE models. A third-party module, called okhttp3, was used to save and manage the SE models. In addition, for SE, CITISEN has two recording modes: prerecording and instant recording. In the prerecording mode, CITISEN records the entire speech signal before processing, whereas in instant recording, CITISEN records and processes the speech signal simultaneously. CITISEN is a standardized SE software with a user interface that can support pretrained SE models trained by various machine learning frameworks, including Keras, PyTorch, and TensorFlow. In addition, the SE models can have different architectures or input acoustic feature formats.

Fig. \ref{fig:implementation} (a) shows the implementation of the SE function in CITISEN, which contains four steps, including audio-recording, pre-processing, speech enhancing, and post-pressing. The details of SE function steps in CITISEN are described as follows.

\subsubsection{Audio recording\label{audio-recording}}
In this step, the application interface is implemented using the Java/Android application programming interface (API) AudioRecord. The AudioRecord saves an audio signal at a sampling rate of 16000 Hz in the signal channel. In the instant recording mode, as AudioRecord processes and analyzes audio data in every 5120 bytes, which is equivalent to 320 sample points per second, the instant recording will approximately have a 20 ms delay. The configuration of the AudioRecord in CITISEN is presented in Table \ref{tab:table1a}.

\begin{table}[!b]
\begin{center}
\label{tab:table1a}
\begin{tabularx}{\columnwidth}{>{\centering}m{2.5cm}|>{\centering\arraybackslash}X}
\toprule
\hline
\textbf{Parameter}&\textbf{Value}\\
\hline
\textbf{Sampling rate}&16000 Hz\\
\textbf{Audio channel}&Mono\\
\textbf{Audio format}&PCM in 16 bits\\
\textbf{Audio buffer size}&5120 bytes\\
\hline
\bottomrule
\end{tabularx}
\caption{AudioRecord configuration in CITISEN. (PCM: pulse-code modulation)}
\end{center}
\end{table}

\subsubsection{Pre-processing}
In this step, CITISEN transfers the data format of the mobile input (byte) to the data format of the SE model input (float). For waveform-based SE models, such as FCN, the preprocessing step transfers the format of time-domain audio signals from bytes to float. For spectral-based SE models, such as DDAE, an additional STFT is required to transfer time-domain signals into frequency-domain signals. CITISEN performs STFT by calling Java/Android API DoubleFFT\_1D in the JTransforms library. By calling this API, a one-dimensional time-domain signal is transferred into a complex matrix. The energy part of the complex matrix is presented as a spectrogram, which is used as the input for spectral-based SE models. The phase part of the complex matrix is reserved and used later to convert the enhanced spectrogram back to the time-domain audio signals.

\subsubsection{Speech-enhancing}

To operate the SE model on mobile devices, the pretrained SE model needs to be packaged into a .pb file. Then, CITISEN calls the Java API, which is built in TensorFlow: TensorFlowInferenceInterface, and passes the assetManager (.pb file) and modelFilename (model Name) to the API. Finally, CITISEN loads the SE model and calculates the enhanced speech signal. This part requires the microprocessor of the mobile device to participate in the calculation, and thus different types of mobile phone models will have different time delays. Currently, we have implemented FCN-based and DDAE-based SE in CITISEN; however, the available SE models can be easily extended by uploading the SE models using the same method.

\subsubsection{Post-processing}
For spectral-based SE models, such as DDAE, the output of the SE model are reconstructed to a time-domain signal. The waveform reconstruction method in CITISEN is the iSTFT, which is implemented with the DoubleFFT\_1D function. For waveform-based SE models, such as FCN, the output is already a time-domain signal and does not require additional conversions. Finally, the data type of the enhanced speech signals is converted to a playable form (from float to byte).

\subsection{MA function}
The MA function of CITISEN aims to adapt the SE model to unknown noises or new speakers, or both. CITISEN provides three different MA modes: noise only (N), speaker only (S), and noise and speaker (N+S). Users can upload a short audio clip of the environment noise or their clean speech signal to the cloud server, and the parameters of the original SE model will be fine-tuned using the uploaded data. Users can then download and use the adapted SE models in CITISEN. Currently, we suggest that users record their referenced target speech signal in a noise-free environment. However, previous studies \cite{zezario2020self, fujimura2021noisy} have shown that some level of noise contained in the referenced target can also lead to an effective reconstruction of the clean waveform in an SE system. The implementation of the MA function is shown in Fig. \ref{fig:implementation} (b).

\subsection{BNC Function}
BNC is a new topic in the field of speech processing. This idea is similar to the changing background of an image or video \cite{seki2000robust}. With the BNC, users can artificially convert the background noise of their speech signal to another specified noise. To use the BNC function, the noises of the target background must be recorded and stored first. Users can record background noises in different environments in real-world scenarios, such as car engine sounds and train stations. Then, users need to select the target background noise before running the BNC function. When running the BNC function, CITISEN removes the original background noise by using SE first and mixes the enhanced speech signal with new background noises by playing them simultaneously. In addition to SE steps, BNC has three additional steps: audio recording (of background noise), noise selection, and audio mixing. Fig. \ref{fig:implementation} (c) illustrates the implementation of the BNC function.

\subsection{CITISEN user interface and usage}
\begin{figure}[htbp!]
{\centering \includegraphics[scale= 0.8]{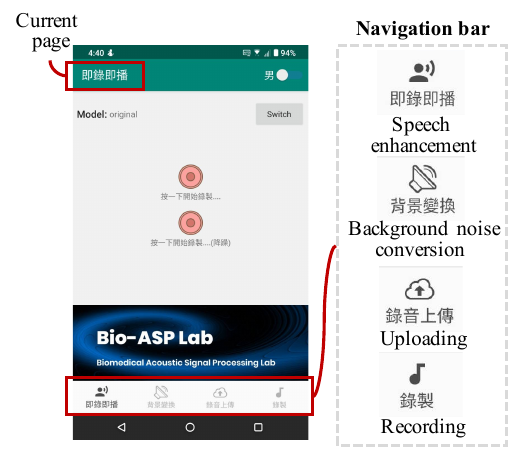}
\caption{Four main pages in CITISEN.}\label{fig:app_main}}
\end{figure}

CITISEN has four pages: “speech enhancement,” “background noise conversion,” “uploading,” and “recording,” as shown in Fig. \ref{fig:app_main}. The page name and navigation buttons are on each page's top left and bottom, respectively.

\subsubsection{Speech enhancement page}
\begin{figure}[htbp!]
{\centering \includegraphics[scale= 0.8]{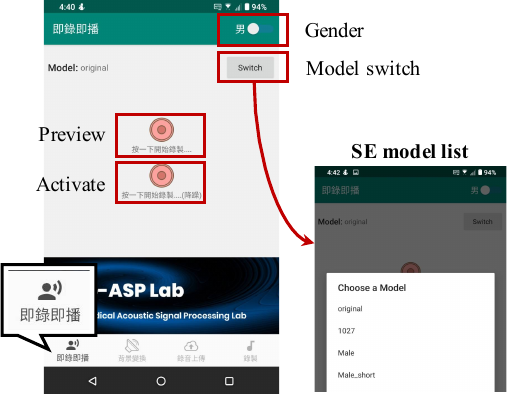}
\caption{Speech enhancement page of CITISEN. The “gender” button on the upper-right corner is used to specify the user's gender. By pressing the “model switch” button, an SE model list will pop up, and users can change the SE model. After pressing the “preview” button, users will hear their original instant recording, and after pressing the “activate” button, users will hear their enhanced instant recording.}\label{fig:app_SE}}
\end{figure}

Fig. \ref{fig:app_SE} shows the “speech enhancement” page. On this page, the user needs to specify the gender by the “gender” button. Because males and females usually have different voice characteristics, knowing the users’ gender can help to improve the performance of SE models. Then, by pressing the “model switch” button, the user can choose different SE models from an SE model list. Currently, CITISEN provides several default SE models trained using our own collected speech datasets. By pressing the “preview” button, users can hear their instant recording without using SE. By pressing the “activate” button, the SE function will be activated, and users can hear their enhanced instant recording.

\subsubsection{Background noise conversion page}

\begin{figure}[htbp!]
{\centering \includegraphics[scale= 0.8]{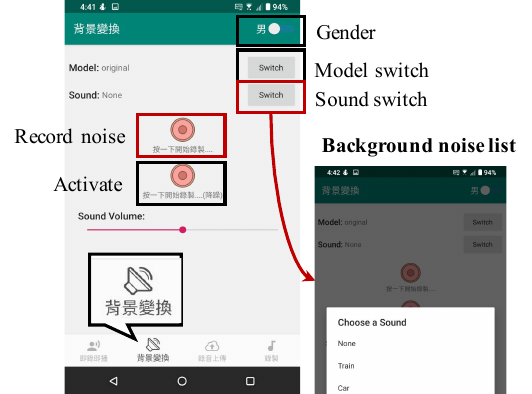}
\caption{Background noise conversion page of CITISEN. By pressing the “sound switch” button, a background noise list will pop up. After pressing the “record noise” button, users can record and save a new noise signal. After pressing the “activate” button, users will hear the enhanced instant recording contaminated with the specified background noise. Note that the “gender” button and the “model switch” button have the same function as those on the “speech enhancement” page.}\label{fig:app_ASC}}
\end{figure}

The “background noise conversion” page of CITISEN is shown in Fig. \ref{fig:app_ASC}. On this page, CITISEN mixes the specified background noise with the enhanced speech signal to generate a new speech signal with the specified background noise. By pressing the “sound switch” button, users can choose the background noise they want to use on the pop-up background noise list. By pressing the “record noise” button, users can record and save a new background noise. In addition, by pressing the “activate” button, users will hear their enhanced instant recording with the specified background noise. Moreover, the “background noise conversion” page has a volume bar, which allows users to adjust the volume of background noise and specify the SNR level of the converted speech signal accordingly.

\subsubsection{Uploading page}

\begin{figure}[htbp!]
{\centering \includegraphics[scale=0.8]{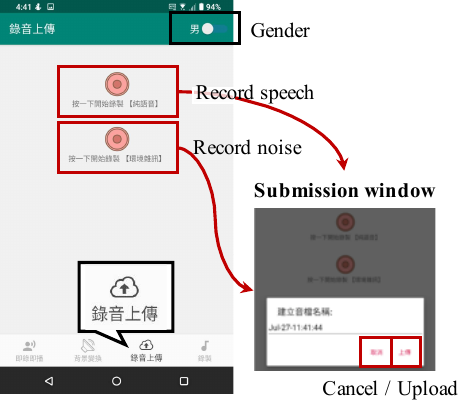}
\caption{Uploading page of CITISEN. After recording a noise or speech signal, CITISEN asks the user to name and save the audio file and upload it to the cloud server.}\label{fig:app_uploading}}
\end{figure}

The “uploading” page is used for uploading the data for the MA function. As CITISEN provides both unknown noise adaptation and new speaker adaptation, there are two file upload buttons: “record speech” and “record noise.” To start the recording, users can simply press one button. After finishing the recording by pressing the button again, CITISEN will pop up a submission window. Users can then name the audio file and upload the recorded audio to the server. After receiving the audio file, the server can adapt the SE model by fine-tuning the original SE model using the recorded audio data. The name of the audio file can also be used to call the adapted SE model, which is later sent from the server to the mobile device and appears on the SE model list on “speech enhancement” and “background noise conversion” pages. Accordingly, users can run the SE and BNC functions using the adapted SE model. The “uploading” page of CITISEN is shown in Fig. \ref{fig:app_uploading}.

\subsubsection{Recording page}

\begin{figure}[htbp!]
{\centering \includegraphics[scale=0.8]{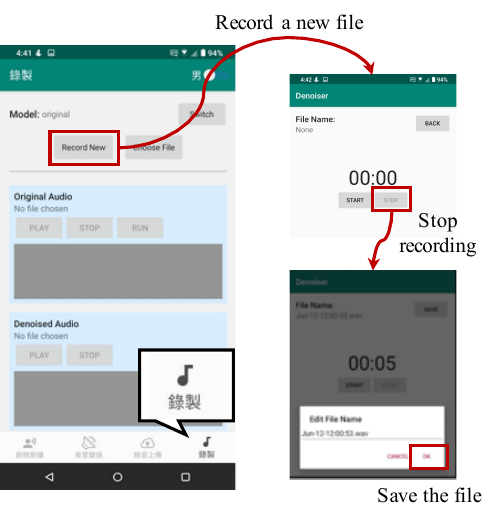}
\caption{Recording page of CITISEN (Part I). 
A new audio file is recorded after pressing the “record new” button. The file can then be named and saved in a pop-up submission window.}\label{fig:app_recording_a}}
\end{figure}

\begin{figure}[htbp!]
{\centering \includegraphics[scale=0.8]{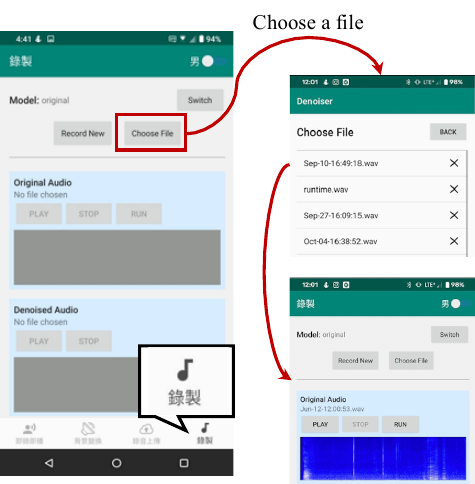}
\caption{Recording page of CITISEN (Part II). By pressing the “choose file” button, users can choose an audio file on a pop-up window.}\label{fig:app_recording_b}}
\end{figure}

\begin{figure}[htbp!]
{\centering \includegraphics[scale=0.8]{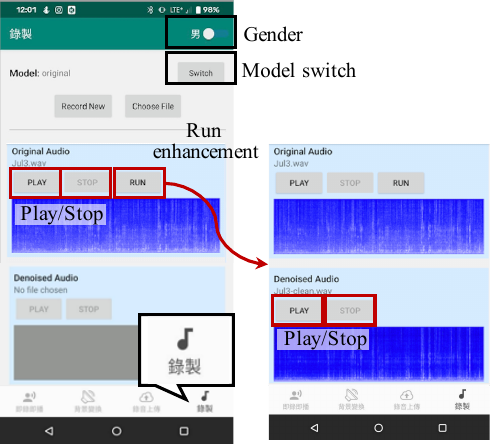}
\caption{Recording page of CITISEN (Part III). Users can choose an SE model type and an SE model by using the “gender” and “model switch” buttons. In addition, users can evaluate the SE results visually and aurally.}\label{fig:app_recording_c}}
\end{figure}

The “recording” page supports prerecording and SE model evaluation. Specifically, on the “recording” page, users can save, playback, and run SE on a saved speech signal. First, users can record new audio by pressing the “record new” button, and CITISEN will redirect to a processing page. After finishing the recording by pressing the “stop” button, users can name and save the record. The workflow is shown in Fig. \ref{fig:app_recording_a}. Then, users can choose an audio file, a model mode, and an SE model with the “choose file,” “gender,” and “model switch” buttons, respectively. Finally, by pressing the “run” button, an enhanced speech signal is generated. Because CITISEN demonstrates both the noisy and enhanced spectrograms, users can visually evaluate the SE results. In addition, users can aurally evaluate the results by pressing the “play” and “stop” buttons to listen to the original and the enhanced speech signals. An illustration showing more details about the “recording” page is shown in Fig. \ref{fig:app_recording_b} and Fig. \ref{fig:app_recording_c}.

\section{EXPERIMENTS}

This section presents the setup, implementation details, and results of the experiments that tested the performance of the SE, MA, and BNC functions. 

\subsection{Experimental setup}
In this study, TMHINT utterances \cite{huang2005development} were used to prepare the training and testing sets, and the utterances were recorded at a 16 kHz sampling rate in a 16-bit format. Notably, the experiments are conducted offline on the cloud platform instead of the mobile platform for several reasons. First, the cloud platform provides a more stable communication and computation environment, which ensures the listening test can go smoothly. Second, because the performance of mobile phones varied too much, choosing one as the representative is hard. Moreover, mobile phones progress so fast that the current best mobile phone might be greatly outperformed by the new mobile phone next year. Finally, evaluating the results on the cloud platform provides the upper bound of these functions and makes the results comparable with other studies.

\subsubsection{SE experiments \label{ sec:SE_experiments}}

In the SE experiments, the training set was prepared using speech utterances from three males and three females. Each speaker read 200 TMHINT utterances in a quiet room, totaling 1200 clean utterances. Each utterance had approximately 3s and contained ten Chinese characters. Noisy utterances were generated by artificially contaminating these 1200 clean training utterances with five randomly sampled noise types from a 100-noise type dataset \cite{hu2004100} at eight different SNR levels ($\pm 1$dB, $\pm 4$dB, $\pm 7$dB, and $\pm 10$dB). Consequently, 48000 noisy-clean pair utterances were obtained. As for the testing set, we used the speech utterance from two other speakers (one male and one female, termed testing speaker in the following discussion), with 120 utterances for each speaker. We generated noisy utterances by artificially contaminating these 120 clean utterances with another set of five noise types (car, sea wave, take-off, train, and song) at two different SNR levels ($0$dB and $5$dB). Notably, the speakers, speech content, and noise types differed between the training and testing sets. The performance of the SE was tested using both subjective listening tests and objective evaluations.

For the listening tests, we recruited 20 participants with a male-to-female ratio of 2 to 3. The group ages were between 20 and 38 years, with a mean age of 21.50 (standard deviation (SD) = 3.97). All participants were native Mandarin speakers with normal hearing to perceive the stimuli during the test. Each participant listened to 80 testing speech signals (40 for 0 dB and 40 for 5 dB) spoken by one male and one female testing speakers. These 80 speech signals had different contents with one of the five assigned background noises (car, sea wave, take-off, train, and song) under four conditions, including original noisy speech signals (without enhancement), enhanced by an MMSE-based SE method, enhanced by a DDAE-based SE model, and enhanced by an FCN-based SE model. These four conditions are denoted as noisy, MMSE, DDAE, and FCN, respectively, in the following discussion. Each participant tested 40 lower- and 40 higher-SNR speech signals. In addition, the subjects were instructed to repeat what they had heard verbally and were allowed to repeat the stimuli once. The character correct rate (CCR), which is calculated by dividing the number of correctly identified characters by the total number of characters, was used to evaluate the intelligibility of speech signals.

For the objective test, we evaluated the results of two more neural-network-based methods, including LSTM-based SE and CRNN-based SE. In the following discussion, the speech signals enhanced by these two methods are denoted as LSTM and CRNN, respectively. PESQ \cite{rix2001perceptual} and STOI \cite{taal2011algorithm} were used as objective evaluation metrics. PESQ was designed to evaluate the quality of the processed speech signal, and the score ranged from -0.5 to 4.5. A higher PESQ score indicates that an enhanced speech signal is closer to the clean speech signal. STOI was designed to compute speech intelligibility, and the scores ranged from 0 to 1. A higher STOI score indicates better speech intelligibility.

\subsubsection{MA experiments}
The performance of the MA function was evaluated under three modes: MA(N), MA(S), and MA(N+S). The training set of the MA experiments was prepared as follows: For MA(N), two new noises (machine beeping and air flowing) from a real hospital scenario were mixed with the same training clean utterances as the SE experiments to form the new noisy-clean speech signal pairs. For MA(S), we mixed 40 clean utterances of the testing speakers in the SE experiments (20 utterances for each speaker) with the same training noises as the SE experiments to form the new noisy-clean speech signal pairs. For MA(S+N), the testing speakers' clean utterances and new noise signals were mixed to form new noisy-clean speech signal pairs. In the SE experiments, the SNRs for performing noisy-training utterances were $\pm 1$ dB, $\pm 4$ dB, $\pm 7$ dB, and $\pm 10$ dB. These training data were then used to fine-tune the pretrained SE model in the SE experiments until the model converged. The testing set of MA experiments had the same testing clean utterances as the SE experiments mixed with machine beeping and air flowing noise at four different SNR levels ($\pm 2$ dB, $0$ dB, and $5$ dB).

Specifically, for MA(N), the training and testing speakers were independent, but the noises came from the same source. For MA(S), the training and testing speakers overlapped, but the training and testing noises were independent. In MA(S+N), the training speakers and testing speakers overlapped, and the noises came from the same source. Note that in every MA experiment, the contents of training speech signals and testing speech signals were different. In addition, the training and testing noises in MA(N) and MA(S+N) were from the same sources but recorded at different times.

\subsubsection{BNC experiments}
Based on our literature survey, there is no standard method for evaluating BNC. Because BNC aims to convert the original background noise into the target background noise, the accuracy (ACC), which is the number of correctly identified types of background noise divided by the total number of questions, was used to evaluate the BNC results. In addition, CCR was used to evaluate the maintenance of clarity and intelligibility of the converted speech signals. The ACC and CCR scores are estimated by both humans and machines. Specifically, we invited human listeners to conduct listening tests. We also trained an ASC model to analyze the ACC and used a pretrained automatic speech recognition (ASR) model to measure the CCR. For human evaluation, the CCR was the ratio of characters that a participant could correctly recognize. For machine evaluation, the CCR was calculated using the Levenshtein distance \cite{levenshtein1966binary} between the predictions of a pretrained ASR system \cite{google_asr} and the ground truth. The details of the listening test and the ASC model are as follows.

\paragraph{Listening test}
We asked the listeners to identify one out of five background noises (car, sea wave, take-off, train, and song) after listening to a converted speech signal. To avoid random guessing, listeners could choose “not clear” if they could not identify the background noise. During the test, participants were asked to repeat what they had heard, select the characters they had heard, and identify the background noise. Forty participants with a male-to-female ratio of 9 to 11 were recruited to participate in this set of listening tests. The group ages were between 14 and 43 years, with a mean age of 25.74 (SD = 8.68). All participants were native Mandarin speakers with normal hearing to perceive the stimuli during the test. 

The stimuli were Mandarin sentences spoken by one male and one female testing speaker. The testing speech signals were either processed using one of three SE methods (i.e., MMSE, DDAE, and FCN) or were not processed (i.e. the clean speech signals). Notably, the enhanced speech signals from SE experiments were used for BNC experiments, which means the original background noises were either car, sea wave, take-off, train, or song. The enhanced speech signals were then contaminated with the car, sea wave, take-off, train, or song noises, making 5 $\times$ 5 possible kinds of BNC conditions. To avoid the fatigue effect, we only tested the results of the 5 dB to 5 dB SNR condition. That is, the SNR of original and converted speech signals are both 5 dB. In total, each participant listened to 80 utterances.

\paragraph{ASC model \label{exp_setting}}

We used the same dataset as the SE experiment to train the ASC model. Specifically, the training and testing utterances were the same as those in the SE experiments described in Section \ref{ sec:SE_experiments}. Thirteen noise types were used for the ASC model. Five of them were test noises used for the SE experiment, including car, sea wave, take-off, train, and song. The remaining eight noises were selected from the training noises of the SE experiment. Each noise segment was cut into two segments with a ratio of one to four. The shorter segment was used for testing, whereas the longer segment was used for training. The training and testing SNR levels were the same as the SE experiment. 

\begin{figure}[hbtp!]
\begin{center}\includegraphics[scale=0.8]{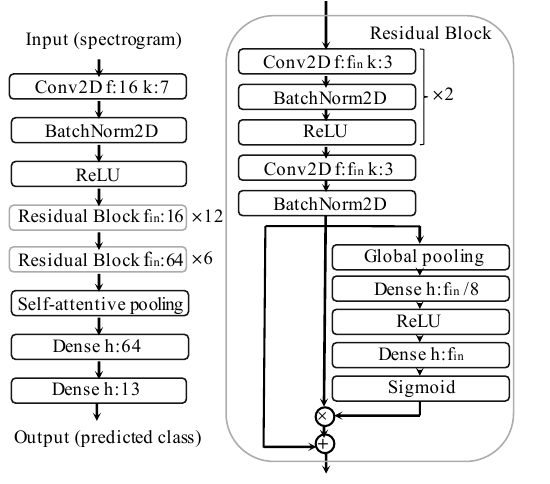}
\caption{Structure of the ASC model. \textbf{Conv2D} represents the 2D convolution layer, where $f$ is the number of filters, and $k$ is the kernel size. The $h$ in the dense layer denotes the size of each output sample.}
\label{fig:ASC_model}
\end{center}
\end{figure}

Fig.~\ref{fig:ASC_model} shows the details of the ASC model, which is based on \cite{chung2020in}. The input of the model is the log1p spectrograms \cite{chuang2020improved}. A Training epoch of 100, batch size of 128, optimizer Adam with a learning rate of 0.0001, and cross-entropy loss were used.

\subsection{Implementation Details of SE models}

This section describes the structures and training details of the neural-network-based SE models. For spectral-based models, including DDAE, LSTM, and CRNN, the parameter settings of the STFT were as follows: the window length was 512, the hop length was 256, and the window type was the Hanning window. Then, the log1p spectrograms \cite{chuang2020improved} were used as the input for the SE models. In inference, the noisy phase was reserved and combined with the enhanced spectral features to reconstruct the time-domain signals.

\subsubsection{FCN}
The FCN consisted of eight convolutional layers, where the filter number and kernel size of each of the first seven layers were 128 and 55, respectively. Batch normalization and the LeakyReLU were used to regularize the output of a hidden layer. The filter number and kernel size in the last layer were 1 and 55, respectively, with the hyperbolic tangent activation function applied to the FCN output. The number of training epochs was set to 60. In addition, batch size 1, optimizer Adam with a learning rate of 0.001, and mean square error (MSE) criteria were used.

\subsubsection{DDAE}

To incorporate contextual information, for each self-defined DDAE layer in this work, five adjacent frames of the input feature vector were concatenated to form the input of the next layer, whereas the output of each layer was a single frame. Also, the ReLU was used to regularize the output layer. The DDAE was composed of three DDAE layers with 257 output units in each layer, followed by a dense layer with single frames as the input and 825 output units, and another dense layer with 257 output units. Finally, the DDAE model had four more DDAE layers with 257 output units. The number of training epochs was 200. In addition, a batch size of 128, an Adam optimizer with a learning rate of 0.0001, and MSE criteria were used.

\subsubsection{LSTM}

The LSTM model used in this evaluation was constructed in the order of three stacked LSTMs and dense layers. Each LSTM layer contained 492 memory cells, and the size of the latest dense layer was 257. The number of training epochs was set to 20. The Adam optimizer with a learning rate of 0.001 and MSE criteria were used. 
 
\subsubsection{CRNN}

The CRNN combines CNN and LSTM to enhance the input raw waveform. The CRNN comprised four convolutional blocks first, where each block was composed of three two-dimensional convolution layers. The ReLu activation function was applied to process the output of each layer. The kernel size for each convolutional layer was three, and the number of channel settings was arranged in the order of 16, 32, 64, and 64. In each block, the stride setting for the output convolutional layer along the speech feature dimension was three, and the setting for the remaining layers was one. Then, the convolutional block was followed by four LSTM layers with 384 memory cells and 257-dimensional dense layers with the ReLu activation function. The input dimensions for the decoder were reshaped from the output of the encoder to 192 ($3\times 64$). In addition, the number of training epochs was 200, the batch size was 128, the optimizer was Adam with a learning rate of 0.0001, and MSE criteria were used.

\subsection{Experimental Results}

In this section, we compare the complexity of the neural-network-based models and then perform a numerical analysis of the SE, MA, and BNC functions. Finally, we present the visualization results of processed speech signals.

\subsubsection{Complexity analyses}
First, we evaluated the complexity of neural-network-based SE models in terms of floating-point operations (FLOPs\footnote{https://github.com/Lyken17/pytorch-OpCounter}) and the number of model parameters. From the results in Table \ref{tab:complex}, we can observe that models with convolutional layers, such as the FCN and CRNN, require higher computational cost in terms of the FLOPs metric. The higher FLOPs imply that these models require more computational loading on hardware resources with similar parameter sizes.

Note that to avoid unstable communication and computation, we conducted experiments offline on a computer. However, we also tested whether the model with the highest FLOPs, the FCN model, could run on CITISEN. The results showed that the FCN model could successfully run on CITISEN.

\begin{table}[!b]
\begin{center}
\begin{tabularx}{\columnwidth}{>{\centering}m{2cm}|>{\centering}m{2.5cm}>{\centering\arraybackslash}X}
\toprule
\hline
&\textbf{FLOPs (M)}&\textbf{\# of parameters (M)}\\
\hline
\textbf{FCN}&10.8&5.4\\
\textbf{DDAE}&2.1&2.1\\
\textbf{LSTM}&5.5&5.5\\
\textbf{C--RNN}&9.5&4.8\\
\hline
\bottomrule
\end{tabularx}
\caption{FLOPs and number of model parameters for FCN, DDAE, LSTM, and CRNN models.}\label{tab:complex}
\end{center}
\end{table}

\subsubsection{SE experiment}\label{SE_experimen}
Table \ref{tab:table0} presents the STOI and PESQ scores of noisy and enhanced speech signals processed using the MMSE, DDAE, FCN, LSTM, and CRNN models. From Table \ref{tab:table0}, all SE methods improved the PESQ scores, and except for MMSE and LSTM, other SE methods increased the performance of STOI. The increased PESQ along with the decreased STOI imply that some SE methods improve the quality, but the produced distortion might affect the intelligibility of a speech signal. The results also show that DDAE, CRNN, and FCN achieved higher scores than MMSE in terms of both STOI and PESQ, whereas FCN provided the highest PESQ and STOI scores among the evaluated methods. The results also demonstrate the effectiveness of using a deep-learning model for the SE task. 

Table \ref{tab:table01} presents the subjective listening test results for noisy and the three SE methods. From the table, it can be observed that MMSE yielded lower CCRs compared to noisy for both 0 dB and 5 dB SNRs, which is consistent with the findings of previous research and the STOI results reported in Table 1. That is, although some SE methods effectively remove background noise, speech intelligibility might be affected. In addition, the SE function is more helpful under low SNR situations, as noisy speech signals maintain high levels of intelligibility under high SNR situations. The one-way analysis of variance and Tukey post-hoc comparisons were applied to demonstrate the significance of improvements for analyzing the SNR-based CCR results of noisy, MMSE, FCN, and DDAE. The evaluations first revealed the significant difference across four SE systems, with $p<0.001$ at 0 dB and 5 dB SNRs. The Tukey post-hoc tests further verified the significant differences for the following SE condition pairs at 0 dB: (FCN, DDAE), (DDAE, MMSE), (FCN, MMSE), and (noisy, MMSE), and at 5 dB: (MMSE, DDAE), (noisy, DDAE), (FCN, DDAE). Notably, the analysis on the scores of FCN and noisy indicated no significant difference, with $p>0.05$ at both 0 dB and 5 dB SNRs. To achieve a significant difference from noisy speech signals to enhanced speech signals, a more advanced SE method performing under lower SNR conditions might be required. 

\begin{figure}[htbp!]
\begin{center}\includegraphics[width=\columnwidth]{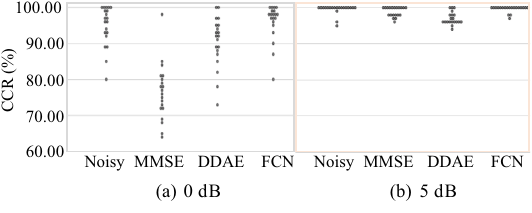}
\caption{The subject-wise CCRs at (a) 0 dB and (b) 5 dB SNR conditions.}\label{fig:subwise}
\end{center}
\end{figure}

In addition to the averaged CCRs for all the participants, Fig. \ref{fig:subwise} (a) and (b) illustrate the subject-wise CCRs at 0 dB and 5 dB, respectively. Each gray circle in the figure represents the CCR score of an individual participant. According to both sub-figures, we can observe a larger CCR variance for MMSE and DDAE than that for FCN and noisy. The results imply the effectiveness of the FCN model in enhancing noisy speech signals with less ambiguous content than that of MMSE and DDAE.

\begin{table}[!b]
\begin{center}
\begin{tabularx}{\columnwidth}{>{\centering}m{2.5cm}|>{\centering}m{2.5cm}>{\centering\arraybackslash}X}
\toprule
\hline
&\textbf{STOI}&\textbf{PESQ}\\
\hline
\textbf{Noisy}&0.6943&1.5188\\
\textbf{MMSE}&0.6497&1.6966\\
\textbf{FCN}&0.7666&2.2518\\
\textbf{DDAE}&0.7260&2.0366\\
\textbf{LSTM}&0.6610&1.6666\\
\textbf{CRNN}&0.7549&2.2144\\
\hline
\bottomrule
\end{tabularx}
\caption{Average STOI and PESQ scores for noisy and three SE methods over 0 and 5 dB SNR conditions. Noisy denotes the results of original noise without performing SE.}
\label{tab:table0}
\end{center}
\end{table}

\begin{table}[!b]
\begin{center}
\begin{tabularx}{\columnwidth}{>{\centering}m{2.5cm}|>{\centering}m{2.5cm}>{\centering\arraybackslash}X}
\toprule
\hline
&\textbf{0 dB}&\textbf{5 dB}\\
\hline
\textbf{Noisy}&0.948&0.995\\
\textbf{MMSE}&0.764&0.989\\
\textbf{DDAE}&0.905&0.970\\
\textbf{FCN}&0.960&0.996\\
\hline
\bottomrule
\end{tabularx}
\caption{Average speech recognition results (CCRs) for noisy and three SE methods at 0 dB and 5 dB SNR conditions.}\label{tab:table01}
\end{center}
\end{table}

\subsubsection{MA experiment}

For the MA experiment, we fine-tuned the FCN model used in the SE experiment and used the original SE results from the FCN model as our baseline. From Table \ref{tab:table1}, it can be seen that SE yielded higher STOI and PESQ scores as compared to noisy, thereby confirming the results in \label{SE_experimen} that SE can improve speech quality and intelligibility over noisy speech signal, although the noise types are unknown and different from those used in the training set.

Next, compared with the baseline (original SE model without MA), all three MA modes achieved higher PESQ and STOI scores. More specifically, MA(N), MA(S), and MA(N+S) yielded noticeable relative improvements of $5.06\%$, $2.94\%$, and $5.84\%$ in terms of STOI, and relative improvements of $12.48\%$, $3.32\%$, and $11.24\%$, in terms of PESQ, respectively, as compared to the baseline. Thus, the results obtained confirmed the effectiveness of the MA function and indicated that intelligibility and quality improvements could be attained by adapting the SE model based on both noise and speaker information. From the experimental results, we also observe that MA(N) achieved a higher PESQ than MA(S) and MA(S + N). One of the possible reasons for this is that the data for MA(N) was more than that for MA(S) and MA(S+N). Specifically, the number of fine-tuned speech signals was 2 $\times$ 1200 $\times$ 8 (new noises $\times$ clean training utterances of the original SE model $\times$ SNRs), 5 $\times$ 40 $\times$ 8 (training noises of original SE model $\times$ clean utterances from new speakers $\times$ SNRs,) and 2 $\times$ 40 $\times$ 8 (new noise $\times$ clean utterances from new speakers $\times$ SNRs) for MA(N), MA(S), and MA(S+N), respectively.

\begin{table}[!b]
\begin{center}
\begin{tabularx}{\columnwidth}{>{\centering}m{2.5cm}|>{\centering}m{2.5cm}>{\centering\arraybackslash}X}
\toprule
\hline
&\textbf{STOI}&\textbf{PESQ}\\
\hline
\textbf{Noisy}&0.7392&1.7976\\
\textbf{Baseline (w/o MA)}&0.7858&2.3888\\
\textbf{MA(N)}&0.8256&2.6870\\
\textbf{MA(S)}&0.8090&2.4681\\
\textbf{MA(N+S)}&0.8317&2.6572\\
\hline
\bottomrule
\end{tabularx}
\caption{Average STOI and PESQ scores for different SE models over -2, 0, 2, and 5 dB SNR conditions. Noisy denotes the results of original noise without performing SE, and baseline denotes the original FCN-based SE results.}\label{tab:table1}
\end{center}
\end{table}

\subsubsection{BNC experiment}
We present human and machine evaluations of the BNC function. Human evaluation was performed by conducting a listening test, whereas the machine evaluation was performed using an ASC model and a pretrained ASR system \cite{google_asr}. We evaluated the BNC using machines for three major reasons. First, recruiting humans to perform the tests is expensive and time-consuming, whereas using a machine to evaluate the performance is relatively inexpensive and efficient. Second, the ASC model has potential in several applications, such as monitoring systems, context-aware mobile devices, and audio search. Third, the machine can assist in human judgment. Therefore, the performance of the ASC model is also important for the BNC function. The details of the ASC model used in this study are described in Section~\ref{exp_setting}. 

\paragraph{Results of human evaluation}

Based on the three SE methods, namely, MMSE, FCN, and DDAE, three sets of converted speech signals were obtained, denoted as BNC(MMSE), BNC(FCN), and BNC(DDAE), respectively. In addition, we included the results of BNC(clean), which is a set of speech signals converted from a silent background. Notably, BNC(clean) represents the upper bound of the BNC results because it was converted from a clean speech signal. From Table~\ref{tab:asc_listening}, we find that BNC(clean) has an ACC of 86.5\%. This result suggests that participants sometimes could not correctly identify the type of background noise, although other types of noise did not contaminate the original speech signal. The 54.9\% ACC of the BNC(MMSE) indicated that the enhanced speech signals of MMSE still contained high noise components that hindered the identification of the new background. However, BNC(FCN) and BNC(DDAE) achieved approximately 80\% of ACC, suggesting that FCN and DDAE can produce enhanced speech signals with low residual noise components for the BNC function. Finally, the high CCR scores of the BNC(FCN) and BNC(DDAE) indicate the maintenance of clarity and intelligibility of the converted speech signals.

\begin{table}[!b]
\begin{center}
\begin{tabularx}{\columnwidth}{>{\centering}m{2.5cm}|>{\centering}m{2.5cm}>{\centering\arraybackslash}X}
\toprule
\hline
&\textbf{CCR}&\textbf{ACC}\\
\hline
\textbf{BNC(clean)}& 0.983 &0.865\\
\textbf{BNC(MMSE)}& 0.946&0.549\\
\textbf{BNC(FCN)}&  0.968 &0.814\\
\textbf{BNC(DDAE)}& 0.949&0.804\\
\hline
\bottomrule
\end{tabularx}
\caption{CCR and ACC scores based on the BNS function in CITISEN.\label{tab:asc_listening}}
\end{center}
\end{table}

\begin{figure}[htbp!]
\begin{center}\includegraphics[width=0.95\columnwidth]{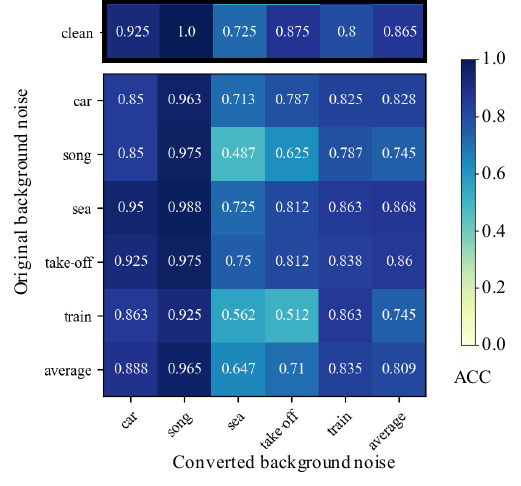}
\caption{Listening test ACC of different kinds of BNC conditions. The results were the average scores of BNC(FCN) and BNC(DDAE).}
\label{fig:listening_asc}
\end{center}
\end{figure}

Fig.~\ref{fig:listening_asc} shows the ACC of different types of BNC conditions. The results were the average scores of BNC(FCN) and BNC(DDAE). We excluded the results of BNC(MMSE) because the ACC of BNC(MMSE) was considerably lower than that of BNC(FCN) and BNC(DDAE), and MMSE performed worse than other SE methods (Table ~\ref{tab:table0}). From the column “sea” and “take-off” in Fig. ~\ref{fig:listening_asc}, we observed that participants were less able to identify the background “sea” and “take-off.” The “sea” and “take-off” backgrounds are less recognizable than the other noises because participants must hear a nearly complete wave or take-off sound to confirm it. Conversely, from the column “song,” we know that participants found it easier to identify the “song” background. This result might be because the “song” background contains music with a human voice, which is considerably different from other background noise. Evidently, the characteristics of the target background significantly affected the identification results. In addition, the original background noise affected the ACC because the noise type usually notably affects the SE performance.

\paragraph{Results of machine evaluation}

Because the BNC function focused on the background noise, the SNR level affected the performance. We make two assumptions about the effect of the SNR level of a speech signal on the performance of the BNC. The first assumption is that a higher original SNR will lead to a better ACC. That is, the target background noise is easier to identify if the converted speech signal is less affected by the original background noise. The second assumption is that a lower converted SNR will result in a better ACC. That is, the target background noise is easier to recognize if it is louder than the speech signals.

\begin{figure}[hbtp!]
\begin{center}\includegraphics[width=1.0\columnwidth]{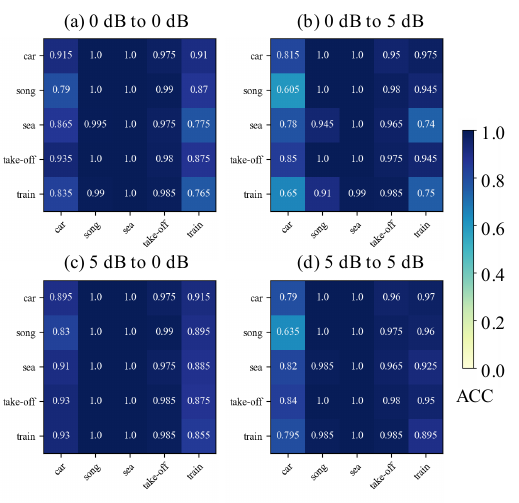}
\caption{ACC at different BNC conditions. Comparing (c) and (d) (bottom row) with (a) and (b) (top row), the results showed that a speech signal with a higher original SNR had a better BNC result. In addition, comparing (a) and (c) (left column) with (b) and (d) (right column), the results indicated that a speech signal with a lower converted SNR had a better BNC result. The results were the average scores of BNC(FCN) and BNC(DDAE).}
\label{fig:asc_machine}
\end{center}
\end{figure}

To test these two assumptions, we conducted four pairs of experiments that converted speech signals with the original SNR level $a$ dB to speech signals with converted SNR level $b$ dB, where $a \in \{0, 5\}$, and $b \in \{0,5\}$. Fig. ~\ref{fig:asc_machine} shows the average ACC of the BNC(DDAE) and BNC(FCN). The figures in the same row and column represent the speech signals with the same original levels and converted SNR levels, respectively. That is, the influences of the original SNR level could be obtained by comparing the figures in different rows, whereas the effects of the converted SNR level could be determined by comparing the figures in different columns. In Fig.~\ref{fig:asc_machine}, we find that speech signals with an original SNR level of 5 dB (bottom row) outperform speech signals with an original SNR of 0 dB (top row), which confirms our first assumption that a speech signal with a higher original SNR has a better BNC result. Subsequently, speech signals with the converted SNR level of 0 dB (left column) performed better than speech signals with the converted SNR level of 5 dB (right column). This result verified our second assumption that a speech signal with a lower converted SNR yields a better BNC result. 

\begin{figure}[hbtp!]
\begin{center}\includegraphics[width=1.0\columnwidth]{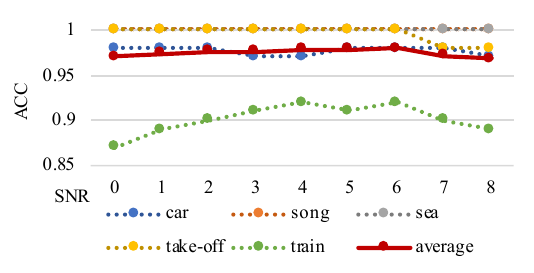}
\caption{ACC vs. SNR of speech signals converted from a silent background. The ACC of “song” and “sea” overlapped in this figure. The results showed that SNR levels did not affect the ACC of BNC.}
\label{fig:asc_machine_clean}
\end{center}
\end{figure}

\begin{figure}[htbp!]
\begin{center}\includegraphics[width=1.0\columnwidth]{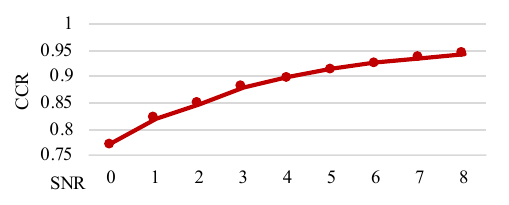}
\caption{CCR vs. SNR of speech signals converted from a silent background. The results showed that the lower the SNR levels, the lower was the CCR.}
\label{fig:asc_machine_clean_ccr}
\end{center}
\end{figure}

We evaluated the ACC of speech signals converted from a silent background (i.e., from a clean speech signal instead of an enhanced speech signal), which is the upper bound of BNC performance. Fig.~\ref{fig:asc_machine_clean} shows the results for different SNR levels. Unlike the results of the previous experiments, the converted SNR level did not affect the ACC of the BNC. None of the background noise conditions indicated that a lower converted SNR would lead to a better ACC. In addition, the average scores remained stable under different SNR levels. One possible reason is that, for enhanced speech signal, a lower converted SNR can suppress the noise that was not removed by the SE models and make the target background easier to identify. Conversely, a lower converted SNR makes no difference for a clean speech signal because it does not contain other noise. Therefore, the target background is easy to recognize despite having a high converted SNR level. Notably, the ASC model achieves high ACC on the “sea” and “take-off” background, whereas the participants of the listening test have a lower identification rate for these two noises. The results suggest that the ASC model has potential to assist human listeners in recognizing noise that they cannot distinguish correctly. Finally, we present the CCR results using a pretrained ASR system \cite{google_asr}. As can be seen in Fig.~\ref{fig:asc_machine_clean_ccr}, the SNR levels significantly affect the CCR. That is, the lower the SNR levels, the lower is the CCR.

Table~\ref{tab:machine_summary} presents a summary of the machine evaluations. For the accuracy of BNC, enhanced speech signals performed worse than clean speech signals but still achieved more than 90\% of accuracy. For the CCR, the performance decreased when using enhanced speech signals instead of clean speech signals. One possible reason is that the ASR system was not trained with the enhanced speech signals; therefore, the prediction of an enhanced speech signal is less accurate. In addition, the CCR is significantly affected by the language model of the pretrained ASR system. Specifically, despite having the same pronunciation, the ASR system might result in the wrong word, leading to a decline in CCR.

\begin{table}[htbp!]
\begin{center}
\begin{tabularx}{\columnwidth}{>{\centering}m{2.cm}|>{\centering}m{1.5cm}>{\centering}m{1.5cm}>{\centering}m{1.5cm}>{\centering\arraybackslash}X}
\toprule
\hline
Source & \multicolumn{2}{c|}{Clean} & \multicolumn{2}{c}{Enhanced (5 dB)} \\ \hline
\multicolumn{1}{c|}{Converted SNR} & 5 & \multicolumn{1}{c|}{0} & 5 & 0 \\ \hline
\multicolumn{1}{c|}{ACC} & 0.978 & \multicolumn{1}{c|}{0.970} & 0.937& 0.953\\
\multicolumn{1}{c|}{CCR} & 0.914 & \multicolumn{1}{c|}{0.771} & 0.756 & 0.605\\  
\hline
\bottomrule
\end{tabularx}
\caption{Summary of machine evaluations. The enhanced signal achieved more than 90\% of ACC but had a great decline in CCR.
}
\label{tab:machine_summary}
\end{center}
\end{table}

\begin{figure}[htbp!]
\begin{center}\includegraphics[scale=0.8]{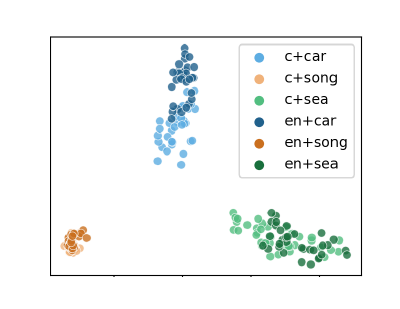}
\caption{Visualization of the ASC embeddings. The original background noise of an enhanced speech signal was either “take-off” or “train.” The embeddings of clean speech signals with new noise were close to those of enhanced speech signals with the same converted noise, which indicates that the BNC function might be used as a data augmentation method for the ASC model when a clean speech signal is unavailable.}
\label{fig:embeddings}
\end{center}
\end{figure}

Subsequently, we used principal component analysis (PCA) \cite{dunteman1989principal} to visualize the embeddings of the ASC models in Fig.~\ref{fig:embeddings}, where the clean and enhanced speech signals with background noise “n” are denoted as “c+n” and “en+n,” respectively. We first found that different noise types were separated, indicating that the ASC model correctly recognized the background noise types. Then, we observed that the embeddings of clean speech signals with a new noise were close to those of enhanced speech signals with the same converted noise. Therefore, the BNC function can serve as a data augmentation method for the ASC model when clean speech signals are unavailable. Specifically, the BNC function can perform data augmentation by generating arbitrary numbers and SNR levels of training speech signals with specific background noise. In addition, the proposed BNC has the potential to open up new and interesting topics that have not yet received sufficient attention. For example, related studies include conversion of a new background noise naturally and the development of an ASC model that can distinguish between artificially converted and naturally recorded background noise.

\subsubsection{Visualization results}
\begin{figure}[htbp!]
{\centering \includegraphics[width=0.95\columnwidth]{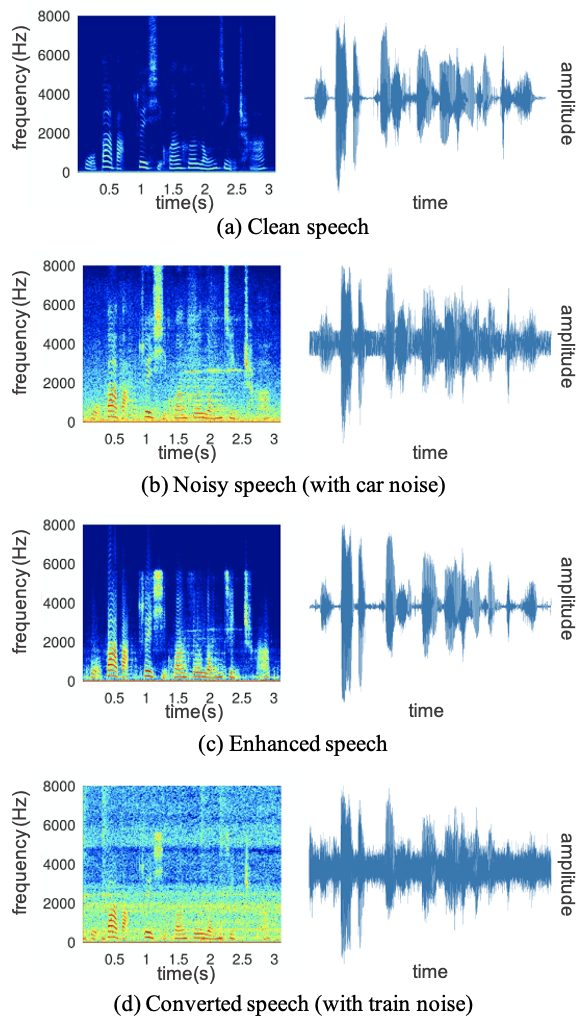}
\caption{CITISEN processed speech signals: (a) clean speech signal, (b) noisy speech signal (with car noise), (c) enhanced speech signal, and (d) converted speech signal (from car noise to train noise). For each sub-figure, the left and right columns show the spectrogram and waveform, respectively.}\label{fig:specwav}}
\end{figure}

Finally, we present the visualization results shown in Fig. \ref{fig:specwav}. Figs. \ref{fig:specwav} (a), (b), (c), and (d) depict the spectrogram and waveform plots of the clean, noisy, enhanced, and BNC speech signals, respectively. For each sub-figure in Fig. \ref{fig:specwav}, the left column depicts the spectrogram, and the right side depicts the associated waveform. Noisy speech signal (b) was produced by contaminating clean speech signal with car noise. Additionally, the BNC speech signal (d), which was produced by mixing the enhanced speech signal (c) with train noise, demonstrates the converted result from car noise to train noise.

The enhanced spectrogram shown in Fig. \ref{fig:specwav} (c) preserves several harmonic clean speech structures when compared with those presented in Figs. \ref{fig:specwav} (a). In addition, when comparing the waveforms in Figs. \ref{fig:specwav} (a), (b), and (c), the enhanced waveform presented in Fig. \ref{fig:specwav} (c) depicts considerably smaller noise components. Both observations demonstrate the effectiveness of SE in reducing noise from noisy input while providing detailed speech structures. The spectrogram shown in Fig. \ref{fig:specwav} (d) clearly illustrates different noise patterns in comparison with those presented in Fig. \ref{fig:specwav} (b) and confirms the effectiveness of BNC.

\section{Conclusion}
In this study, we presented a speech signal processing mobile application called CITISEN. The contributions of CITISEN are as follows: (1) CITISEN was developed as a standardized SE tool with a user interface for performing SE on an prerecording or instant recording. In addition, experimental results confirmed the SE function of providing improved STOI and PESQ scores. (2) CITISEN has an MA function that allows users to adapt the SE models in terms of personalized testing conditions, and the MA function was proven to provide notable STOI and PESQ improvements as compared to the results without MA. (3) CITISEN provides a BNC function that converts the background noise of a speech signal into another noise. Notably, the BNC function is a novel concept for SE techniques and was implemented in mobile devices for the first time. The listening test results indicated that the BNC function could convert the background noise while maintaining the clarity and intelligibility of the converted speech signals. In addition, machine evaluation experiments showed that the ASC embeddings of clean speech signals with a new noise were close to those of enhanced speech signals with the same converted noise. Therefore, the BNC function can serve as a data augmentation method for the ASC model in the condition that clean speech signals are unavailable. (4) By simply replacing the settings with the associated model, CITISEN can run with other SE models that were not tested in this study. Therefore, CITISEN provides a suitable platform for evaluating deep-learning-based SE models and effectively reduces the development interval for converting deep-learning models to industrial applications.

\bibliographystyle{ieeetr}
\bibliography{ref}

\begin{IEEEbiography}[{\includegraphics[width=1in,height=1.25in,clip,keepaspectratio]{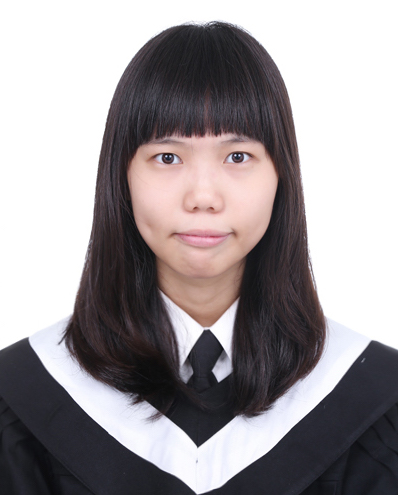}}] {Yu-Wen Chen} received the B.S. degree in electrical engineering from National Cheng Kung University, Tainan, Taiwan, in 2017, and the M.S. degree in electrical engineering from National Taiwan University, Taipei, Taiwan, in 2019. She is currently a research assistant at Academia Sinica, Taipei, Taiwan. Her research interests include speech processing, human-computer interaction, multimodal learning, and machine learning.

\end{IEEEbiography}


\begin{IEEEbiography}[{\includegraphics[width=1in,height=1.25in,clip,keepaspectratio]{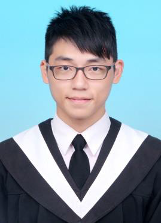}}] {Kuo-Hsuan Hung} received the B.S. and M.S. degrees from the National Chiao Tung University and National Central University, Taiwan, in 2015 and 2017, respectively. He is currently working toward the Ph.D. degree with the Department of Biomedical Engineering, National Taiwan University. He is a Research Assistant with the Research Center for Information Technology Innovation, Academia Sinica, Taipei, Taiwan. His research interests include biomedical signal processing, noise reduction, speaker recognition, and deep learning.
\end{IEEEbiography}


\begin{IEEEbiography}[{\includegraphics[width=1in,height=1.25in,clip,keepaspectratio]{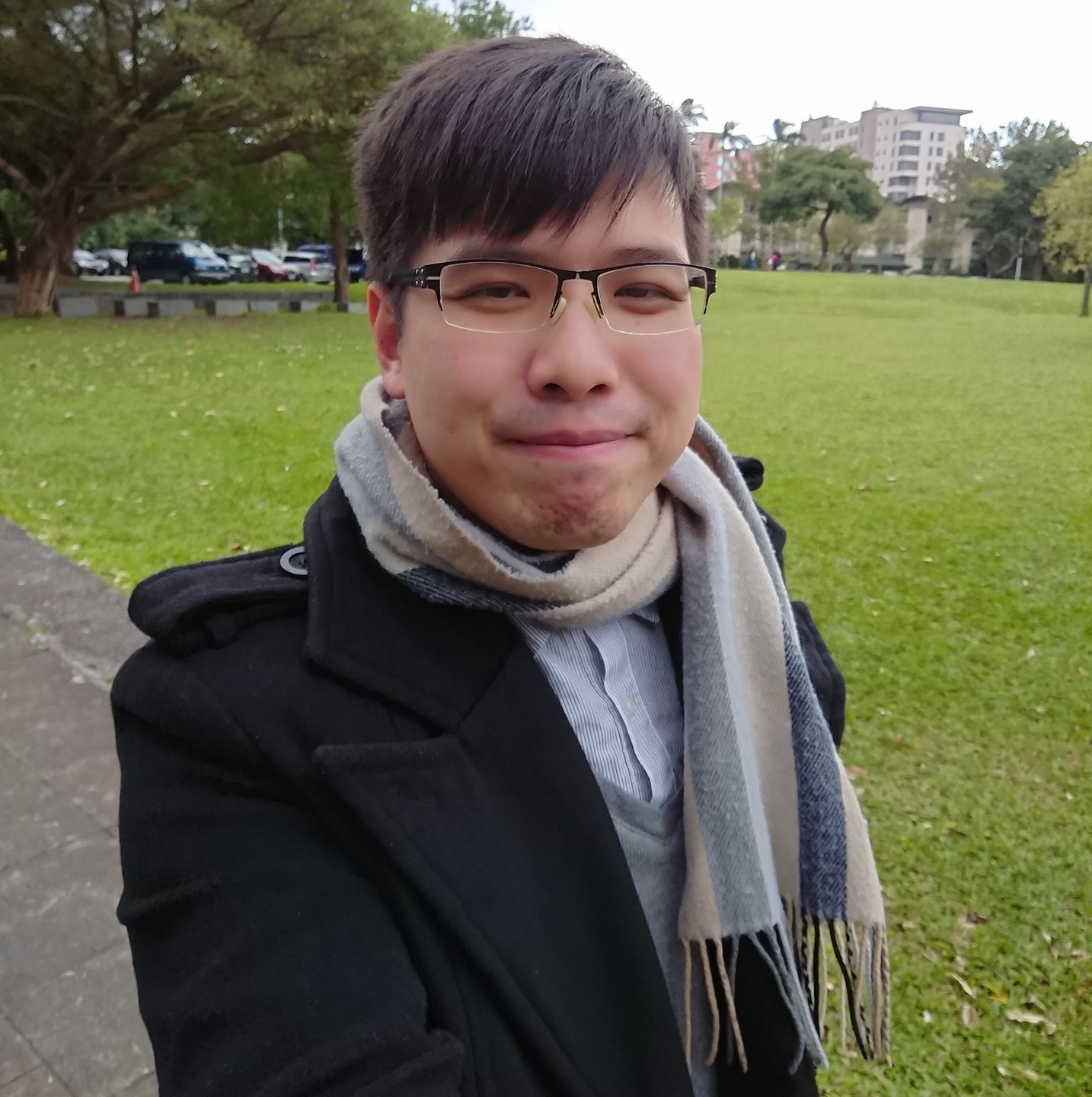}}] {You-Jin Li} received the Ph.D. degree in Education from University of Bath, Bath, UK., in 2020. Her expertise is in the areas of Social and Personality Psychology as well as Sport and Exercise Psychology. She is currently a Postdoctoral Research Fellow with the Research Center for Information Technology Innovation, Academia Sinica, Taipei, Taiwan. Her research interests include psychometric instrument development and testing, parenting education, attachment relationships and psychological well/ill-being.
\end{IEEEbiography}


\begin{IEEEbiography}[{\includegraphics[width=1in,height=1.25in,clip,keepaspectratio]{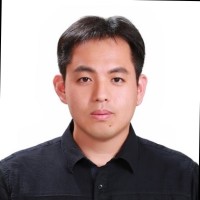}}] {Alexander C. Kang} received the B.S. degrees in electrical engineering from University of California, Los Angeles, CA, USA , in 2009, and the M.S. degree in telecommunication from University of Maryland, College Park, MD, USA, in 2014. From 2014 to 2019, he was a software engineer working in silicon valley, California, USA, where he engaged in artificial intelligence related development. He currently works as a Research Assistant in the Research Center for Information Technology Innovation, Academia Sinica, Taipei. His research interests cover signal processing, speech enhancement, voice conversion, and deep learning.
\end{IEEEbiography}


\begin{IEEEbiography}[{\includegraphics[width=1in,height=1.25in,clip,keepaspectratio]{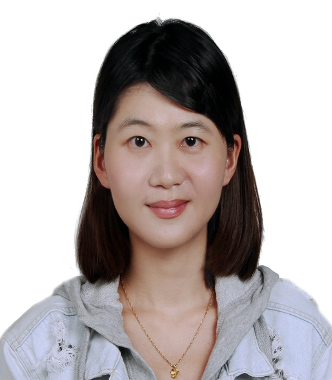}}] {Ya-Hsin Lai} received the Ph.D. degree in Education from University of Bath, Bath, UK., in 2020. From 2020 to 2021, she was a Postdoctoral Research Fellow with the Research Center for Information Technology Innovation, Academia Sinica, Taipei, Taiwan, where she engaged in a variety of research in automatic speech recognition, speech enhancement, and musical intervention for dyslexic children. She is currently an Assistant Professor at the Master Program of Youth and Child Welfare, Chinese Culture University, Taipei, Taiwan. Her research interests include psychometric instrument development and testing, parenting education, attachment relationships as well as child and youth wellness.
\end{IEEEbiography}


\begin{IEEEbiography}[{\includegraphics[width=1in,height=1.25in,clip,keepaspectratio]{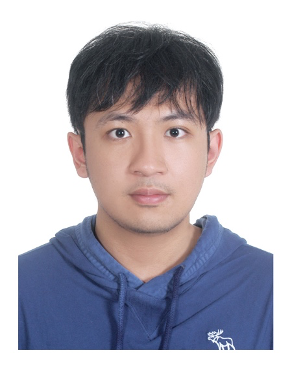}}] {Kai-Chun Liu} received the M.S. and Ph.D. degree in biomedical engineering from National Yang-Ming University, Taipei, Taiwan, in 2015 and 2019. He is currently a Postdoctoral Research Fellow with the Research Center for Information Technology Innovation, Academia Sinica, Taipei. 
His research interests include pervasive healthcare, wearable computing, machine learning and biosignal processing.
\end{IEEEbiography}


\begin{IEEEbiography}[{\includegraphics[width=1in,height=1.25in,clip,keepaspectratio]{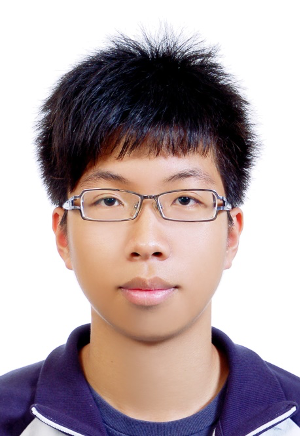}}] {Szu-Wei Fu} received the M.S. and Ph.D. degrees in Graduate Institute of Communication Engineering and Department of Computer Science and Information Engineering from National Taiwan University, Taipei, Taiwan, in 2014 and 2020, respectively. He is currently a Postdoctoral Research Fellow with the Research Center for Information Technology Innovation, Academia Sinica, Taipei. His research interests include speech processing, speech enhancement, machine learning and deep learning.
\end{IEEEbiography}


\begin{IEEEbiography}[{\includegraphics[width=1in,height=1.25in,clip,keepaspectratio]{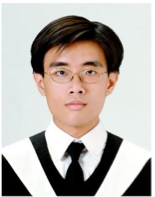}}] {Syu-Siang Wang} received the B.S. degree from the Department of Electrical Engineering, National Changhua University of Education, Changhua, in 2008, the M.S. degree from the Department of Electrical Engineering, National Chi Nan University, in 2010, and the Ph.D. degree from the Graduate Institute of Communication Engineering, National Taiwan University, in 2018. He was a Research Assistant with the Yu Tsao in Research Center for Information Technology Innovation, Academia Sinica, where he was involved in robust speech feature extraction and speech enhancement. He is currently a Research Assistant with Yuan Ze University, Taiwan. His research interests include speech recognition, speech enhancement, audio coding, biosignal processing, and deep neural networks.
\end{IEEEbiography}


\begin{IEEEbiography}[{\includegraphics[width=1in,height=1.25in,clip,keepaspectratio]{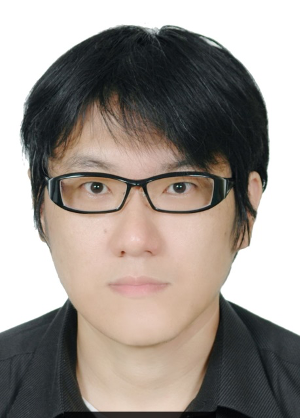}}] {Yu Tsao} (Senior Member, IEEE) received the B.S. and M.S. degrees in electrical engineering from National Taiwan University, Taipei, Taiwan, in 1999 and 2001, respectively, and the Ph.D. degree in electrical and computer engineering from the Georgia Institute of Technology, Atlanta, GA, USA, in 2008. From 2009 to 2011, he was a Researcher with the National Institute of Information and Communications Technology, Tokyo, Japan, where he engaged in research and product development in automatic speech recognition for multilingual speech-to-speech translation. He is currently a Research Fellow (Professor) and Deputy Director with the Research Center for Information Technology Innovation, Academia Sinica, Taipei, Taiwan. He is also a Jointly Appointed Professor with the Department of Electrical Engineering at Chung Yuan Christian University, Taoyuan City, Taiwan. His research interests include assistive oral communication technologies, audio coding, and bio-signal processing. He is currently an Associate Editor for the IEEE/ACM TRANSACTIONS ON AUDIO, SPEECH, AND LANGUAGE PROCESSING and IEEE SIGNAL PROCESSING LETTERS. He received the Academia Sinica Career Development Award in 2017, National Innovation Awards in 2018-2021, Future Tech Breakthrough Award 2019, and Outstanding Elite Award, Chung Hwa Rotary Educational Foundation 2019–2020. He is a co-author of a paper that received the 2021 IEEE Signal Processing Society (SPS), Young Author, Best Paper Award.
\end{IEEEbiography}


\EOD

\end{document}